\documentclass[journal]{IEEEtran}
\ifCLASSINFOpdf
  \usepackage[pdftex]{graphicx}
  \graphicspath{{../pdf/}{../jpeg/}}
  \DeclareGraphicsExtensions{.pdf,.jpeg,.png}
\else
  \usepackage[dvips]{graphicx}
  \graphicspath{{../eps/}}
  \DeclareGraphicsExtensions{.eps}
\fi
\usepackage{caption}
\usepackage{cite}
\usepackage{amsmath,amssymb}
\usepackage{hyperref}
\usepackage{booktabs}
\usepackage{caption}
\usepackage{tabularx}
\usepackage{array}
\usepackage{xcolor}
\usepackage{ragged2e}
\begin{document}

\title{Depression Detection and Analysis using Large Language Models
on Textual and Audio-Visual Modalities}

\author{\IEEEauthorblockN{,
Chayan Tank\IEEEauthorrefmark{1},
Sarthak Pol\IEEEauthorrefmark{2}, 
Vinayak Katoch\IEEEauthorrefmark{3}, 
Shaina Mehta\IEEEauthorrefmark{4}
Avinash Anand\IEEEauthorrefmark{5} and 
Rajiv Ratn Shah\IEEEauthorrefmark{6}
}

\IEEEauthorblockA{\IEEEauthorrefmark{1} \IEEEauthorrefmark{2} \IEEEauthorrefmark{3} \IEEEauthorrefmark{4} \IEEEauthorrefmark{5} \IEEEauthorrefmark{6}} {Indraprastha Institute of Information Technology, Delhi, India.}

\IEEEauthorblockA{ 
\IEEEauthorrefmark{1}chayan23030@iiitd.ac.in 
\IEEEauthorrefmark{2}sarthak23082@iiitd.ac.in
\IEEEauthorrefmark{3}vinayak23105@iiitd.ac.in
\IEEEauthorrefmark{4}shaina23139@iiitd.ac.in
\IEEEauthorrefmark{5}avinasha@iiitd.ac.in 
\IEEEauthorrefmark{6}rajivratn@iiitd.ac.in}
}

\maketitle

\begin{abstract}
Clinical depression, also known as Major depressive disorder (MDD), is a prevalent psychological disorder affecting the general population worldwide. Depression has proven to be a significant public health issue, profoundly affecting the psychological well-being of individuals. If it remains undiagnosed, depression can lead to severe health issues, which can manifest physically and even lead to suicide. Generally, Diagnosing depression or any other mental disorder involves conducting semi-structured interviews alongside supplementary questionnaires, including variants of the Patient Health Questionnaire (PHQ) by Clinicians and mental health professionals. This approach places significant reliance on the experience and judgment of trained physicians, making the diagnosis susceptible to personal biases.
Given that the underlying mechanisms causing depression are still being actively researched, physicians often face challenges in diagnosing and treating the condition, particularly in its early stages of clinical presentation. Recently, significant strides have been made in Artificial neural computing to solve problems involving text, image, and speech in various domains. Our analysis has aimed to leverage these state-of-the-art (SOTA) models in our experiments to achieve optimal outcomes leveraging multiple modalities. The experiments were performed on the Extended Distress Analysis Interview Corpus Wizard of Oz dataset (E-DAIC) corpus presented in the Audio/Visual Emotion Challenge (AVEC) 2019 Challenge. The proposed solutions demonstrate better results achieved by Proprietary and Open-source Large Language Models (LLMs), which achieved a Root Mean Square Error (RMSE) score of 3.98 on Textual Modality, beating the AVEC 2019 challenge baseline results and current SOTA regression analysis architectures. Additionally, the proposed solution achieved an accuracy of 71.43\% in the classification task. The paper also includes a novel audio-visual multi-modal network that predicts PHQ-8 scores with an RMSE of 6.51. The paper also discussed the potential limitations of the dataset and how they can be overcome. 

\end{abstract}

\begin{IEEEkeywords}
Depression Detection and Severity Analysis, Multi-modal Analysis, Deep Learning, Large Language Models, OpenAI Whisper, RoBERTa, BiLSTM, DepRoBERTa, OpenAI GPT 3.5, OpenAI GPT 4, LLAMA 3 8B Instruct.
\end{IEEEkeywords}

\IEEEpeerreviewmaketitle

\section{Introduction}

\begin{figure*}[h]
  \centering
  \includegraphics[width=0.8\textwidth]{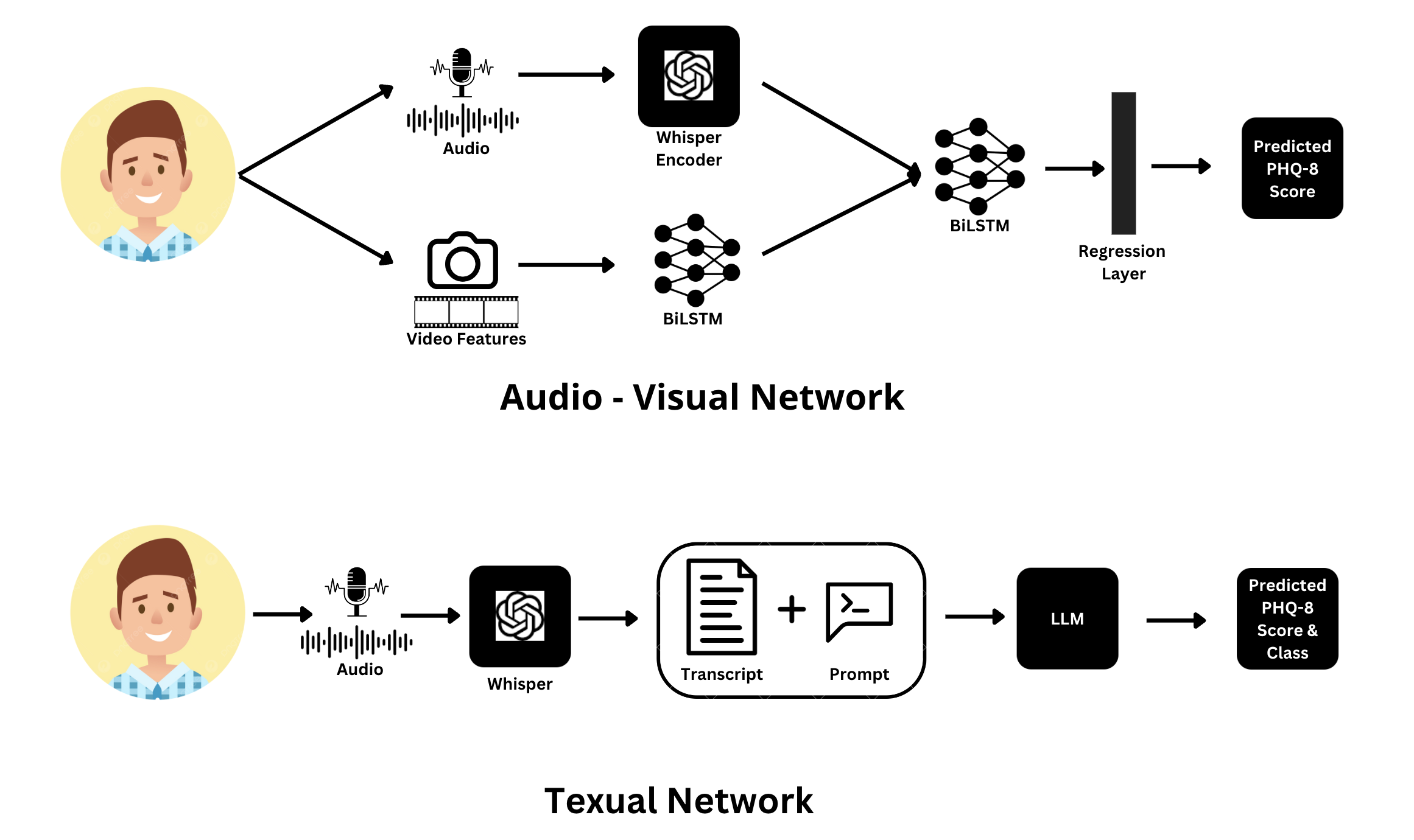}
  \caption {\small The Proposed Textual Network \& Audio - visual network, which predicts the PHQ-8 scores of patients using their Audio, visual and textual clues. In the Textual Network, we have used whisper to extract the transcripts from audio and input them to LLMs along with prompts for PHQ-8 Score and Class prediction. In the Audio-Visual network, we use a Whisper + BiLSTM-based network, which outputs the Predicted PHQ-8 scores}
  \label{fig:Network}
\end{figure*}
\IEEEPARstart{D}{epression} is a widespread and debilitating psychiatric disorder experienced by people around the world \cite{b1}. It is marked by a continuous low mood and a diminished interest or pleasure in everyday activities. The effects of this disorder go well beyond emotional suffering, affecting all areas of a person's life, including their personal health and societal productivity. The prevalence of depression is alarmingly high, with estimates suggesting that it affects over 3.8\% of the global population \cite{b2}. This statistic translates to about 280 million people grappling with this condition, spanning various demographics and geographies. Depression manifests uniquely across various age groups and genders, with around 5\% of adults affected by the condition—comprising 4\% of men and 6\% of women. This indicates that women are more prone to depression than men. Among older adults, particularly those over 60 years old, the prevalence increases to approximately 5.7\%.
These numbers underline the significant impact of depression on public health and the critical need for effective management strategies. Depression can be triggered by a range of factors, both biological and environmental. Individuals who have undergone traumatic experiences, such as abuse or severe loss, are particularly susceptible to developing depression. Additionally, certain biological factors, including genetics and changes in brain chemistry, play crucial roles in the onset and progression of the disorder. This complex interplay of factors makes the detection and treatment of depression a challenging endeavour.\cite{b1}

One of the primary challenges in managing depression effectively is its diagnosis. Traditional methods for diagnosing depression primarily rely on subjective assessments, such as patient-reported questionnaires like the PHQ. These tools depend on the individuals' ability to report their feelings and behaviours accurately, a process often complicated by the nature of depression itself, which can distort self-perception and awareness. The subjective nature of these diagnostic tools can lead to considerable variability in diagnosing depression, with significant implications for treatment. Misdiagnosis or delayed diagnosis can prevent individuals from receiving the appropriate care and potentially aggravate the severity of the condition. Moreover, the lack of straightforward, objective diagnostic tests means that clinicians must rely on regular screenings to monitor the severity of depression in patients, a process that is not only time-consuming but also fraught with the potential for inconsistency.

Given these challenges, there is increasing interest in affective computing, especially in utilising behavioural clues to help detect and assess depression. Quantifiable data from behavioural markers like facial expressions, speech patterns, and changes in physical activity levels can be analysed objectively. These markers can offer insights into an individual's emotional state, potentially helping clinicians detect depression more reliably and at an earlier stage. This emerging approach is part of a broader trend towards multi-modal frameworks in psychiatric assessment, where data from different sources are integrated to form a more complete and precise picture of a patient's mental health. By combining traditional psychological assessments with cutting-edge technology in behaviour analysis, researchers hope to develop more robust methods for not only diagnosing depression but also assessing its severity.

In conclusion, Depression continues to be a significant global health issue due to its high prevalence, profound impact on quality of life, and the complexities involved in its diagnosis and treatment. Developing new tools that offer more objective and reliable assessments of depression is essential. This paper introduces a multi-modal architecture and analysis designed to utilise behavioural clues for more effective depression detection, marking a significant advancement in mental health diagnostics.

\section{Literature Review}

The AVEC workshop was started in 2011 and ended in 2019. Initially, the challenge focuses on multi-modal approaches for sentiment analysis \cite{b25}. Later, specific challenges related to mental health diagnosis using Artificial Intelligence were introduced in the Depression Detection task introduced in the AVEC 2013 challenge \cite{b26}. In AVEC 2016 \cite{b27}, and 2017 \cite{b28} and AVEC 2019 challenge \cite{b5}, Distress Analysis Interview Corpus (DAIC) \cite{b10} \cite{b11}, and E-DAIC dataset was used for depression detection and analysis respectively. The baseline system of depression detection and analysis of these AVEC challenges and their results for their multi-modal frameworks combining audio-video on the test set is given in Table\ref{tab:baseline_system}.

\begin{table}[ht]
\centering
\renewcommand{\arraystretch}{1.5} % Adjust the row height
\begin{tabular}{|c|c|c|c|}
\hline
AVEC Challenge & Task & Dataset & Results \\
\hline
2016 \cite{b27} & Binary Classification & DAIC & F1 - 0.583 \\
\hline
2017 \cite{b28} & PHQ-8 Score Prediction & DAIC & RMSE - 7.05 \\
% & & & MAE - 5.66 \\
\hline
2019 \cite{b5} & PHQ-8 Score Prediction & E-DAIC & RMSE - 6.37 \\
% & & & CCC - 0.111 \\
\hline
\end{tabular}
\caption{\small Baseline System Depression Detection and Analysis in AVEC 2016, 2017 and 2019 Challenges}
\label{tab:baseline_system}
\end{table}

\subsection{AVEC 2019 Challenge}

From AVEC 2019 challenge \cite{b5} published papers, some of the exciting approaches we found based on machine learning and deep learning-based techniques were mentioned in \cite{b29}, \cite{b6}, \cite{b20}, \cite{b12}, and \cite{b4}. Firstly, Zhang et al. \cite{b29}, who segregated the voices of Ellie and the patient from the audio recordings by using the timestamps given in the transcripts and extracted Extended Geneva Minimalistic Acoustic Parameter Set (eGeMAPS) features using DigiVoice featurisation pipeline and based features using Collaborative Voice Analysis Repository (COVAREP) toolkit and applied Logistic Regression with L1 regularisation and Random Forest Algorithm with hyperparameter tuning. They achieved the RMSE and Mean Absolute Error (MAE) of 6.78 and 5.77 on audio modality, which is lower than the baseline results of the AVEC 2019 challenge \cite{b5}. Another research is proposed by Fan et al. \cite{b6}, created a multi-scale temporal dilated Convolutional Neural Network (CNN) for depression severity analysis on audio and textual features of the E-DAIC dataset and achieved the RMSE of 5.07 and 5.91, MAE of 4.06 and 4.39 and Concordance Correlation Coefficient (CCC) of 0.466 and 0.430 on validation and test data, respectively. Makiuchi et al. \cite{b20} applied a multi-modal fusion approach which combines audio and textual features for depression assessment by processing audio features from a pre-trained VGG-16 network through a Gated Convolutional Neural Network (GCNN) followed by a Long Short Term Memory (LSTM) layer and textual features obtained from Bidirectional Encoder Representations from Transformers (BERT) are processed using a CNN followed by an LSTM layer. Then, all the features were fused using a fully connected layer and yielded a CCC score of 0.696 and 0.403 on the development and test sets, respectively.

\vspace{1\baselineskip}

Yin et al. \cite{b12} suggested a multi-modal architecture for depression assessment using a hierarchical recurrent neural network which incorporates two hierarchies of Bidirectional Long Short-Term Memory (Bi-LSTM) memories for multi-modal fusion of data and applied the adaptive sample weighting mechanism to training data. On the validation and test sets, they achieved RMSE and CCC values of 4.94 and 0.402 and 5.50 and 0.442. Finally, Ray et al. \cite{b4} introduced a multi-level attention network consisting of only Bi-LSTM and attention mechanism for the depression severity analysis task. They achieved the RMSE, MAE and CCC of 4.37, 4.02 and 0.67 on textual modality, winning the AVEC 2019 challenge \cite{b5}. However, the models she proposed based on audio, video, and multi-modality beat the baseline results of the challenge.

\subsection{Post AVEC 2019 Challenge}
After the AVEC challenge, several researchers have also proposed various machine learning and deep learning-based architectures capable of handling single as well as multi-modality on several datasets, mainly focusing on AVEC 2016, 2017 and 2019 challenges \cite{b27}, \cite{b28}, and \cite{b5} datasets. Firstly, Zhang et al. \cite{b7} introduced a multi-modal framework for depression assessment by processing audio and video features using Multi-modal Deep Denoising Auto-Encoder (DDAE) and encoding them into fisher vectors. Relevant features are selected using the tree-based model. They used paragraph vector (doc-to-vec) models to extract textual-level features. Finally, the Multi-task Deep Neural Network is trained on these features plus the ResNet features and achieves a CCC of 0.560 in textual features, Mean Squared Error (MSE) of 20.06 and F1 Score of 91.7 per cent on multi-modal settings, accuracy of 89.3 per cent on audio-visual modalities, on the development set of EDAIC dataset \cite{b5}. Another research is proposed by Jo et al. \cite{b9} who introduced a four-stream depression detection model that uses an ensemble of Bi-LSTM and CNN to analyse audio and textual elements. They trained the model using the DAIC \cite{b10} \cite{b11}, and E-DAIC \cite{b5} datasets and obtained F1 scores of 97 per cent and 99 per cent, Precision of 97 per cent and 100 per cent, and Recall of 97 per cent and 98 per cent, respectively. Sun et al. \cite{b13} proposed the multi-modal framework for depression analysis known as Cube MLP, which comprises three separate MLP units, each with two affine transformations, and each unit performs separate operations that are sequential mixing, modality mixing, and channel mixing on the data. They achieved an MAE and CCC of 4.37 and 0.583, respectively, on the E-DAIC dataset \cite{b5}. 

\vspace{1\baselineskip}

Yuan et al. \cite{b16} extracted textual features using sentence-level vector (sent2vec) encoders, the Universal Sentence Encoder, and Bi-LSTM and extracted audio and visual features using a combination of the PCA and Midmax algorithms. Then, they applied the Multi-modal Multi-order Factor Fusion (MMFF) algorithm on all the features, achieving RMSE, MAE and CCC of 4.91, 3.98, and 0.676, respectively, on a test set of the E-DAIC dataset \cite{b5}. Saggu et al. \cite{b21} proposed DepressNet as a novel multi-modal machine learning framework for depression detection, which uses a Bi-LSTM layer network with an attention mechanism. This approach yields an RMSE and CCC of 4.32 and 5.36, and 0.662 and 0.457 development and test set of the E-DAIC dataset \cite{b5} respectively. Wang et al. \cite{b33} processed head pose and Action Units (AUs) as features and audio-based features from the COVAREP toolkit using CNN and LSTM in series and converted them into a feature matrix. The textual features are extracted using Sentence BERT (s-BERT) and passed through the network consisting of CNN and Bi-LSTM layers. Then, all the features are passed through Multi-modal Multi-Utterance-Bimodal Attention Networks, followed by an attention mechanism, two LSTM and two dense layers. They trained their network on E-DAIC \cite{b5} and DAIC \cite{b10} \cite{b11} dataset and achieved the RMSE and MAE of 4.03 and 3.05, respectively, on the development set.

\vspace{1\baselineskip}

Sun et al. \cite{b34} proposed a multi-modal architecture called 'Tensorformer', which allows all the modalities to exchange all the relevant information simultaneously. They trained the network on the E-DAIC \cite{b5} dataset and achieved RMSE and CCC scores of 4.31 and 0.491, respectively, on the test set, beating the SOTA model of AVEC 2019 DDS challenge winner \cite{b4} in terms of RMSE only. Mao et al. \cite{b32} categorised the severity of depression based on the PHQ-8 Score into five categories that are healthy, mild, moderate, moderately severe, and severe. They used COVAREP features of the audio and Bi-LSTM and time-distributed CNN, and for textual modality, they used global vectors for word representation (GloVe) embeddings fed into the Bi-LSTM network. Finally, the fusion of modalities is performed using self-attention, dense layers, and majority voting. They achieved the F1 score of 95.80 per cent on patient-level depression detection on the DAIC dataset \cite{b10} and \cite{b11}. Teng et al. \cite{b19} proposed a study Integrating AVEC 2019 \cite{b5} and CMU-MOSEI \cite{b15} datasets. Their research adopts a multi-modal, multi-task learning approach to enhance depression detection accuracy. By leveraging emotional data, their method significantly boosts precision, achieving a CCC and MAE of 0.466 and 5.21 on a test set of the E-DAIC dataset.

\vspace{1\baselineskip}

Li et al. \cite{b22} proposed the Flexible Parallel Transformer (FPT) model, which integrates video and audio descriptors. Tested on the E-DAIC dataset, the FPT model achieved an RMSE of 4.80, an MAE of 4.58, and a depression classification accuracy of 0.79 at a PHQ-8 threshold of 10, outperforming fewer complex models and proving the effectiveness of multi-modal approaches. Gómez et al. \cite{b42} proposed a multi-level temporal model analogous to multi-modal transformer architecture for depression detection by processing audio-visual embeddings, facial and body landmarks, and feeding into the transformer-based encoder. This approach achieved the precision, recall and F1-scores of 74 per cent, 84 per cent and 78 per cent, respectively, on DAIC \cite{b10} \cite{b11} dataset and 59 per cent, 58 per cent and 56 per cent, respectively, on E-DAIC dataset respectively. Steijn et al. \cite{b18} converted the audio into text using the Google ASR toolkit and extracted textual features using s-BERT and Linguistic Inquiry and Word Count (LIWC) toolkit and handcrafted features such as emotion, speech rate, etc., from the audio files. They also used the Kernel Extreme Learning Machine (KELM) algorithm for single multi-task regression and single-task classification purposes and the K-Fold cross-validation technique. They achieved RMSE and CCC scores of 6.06 and 0.62 when performing a summation of multi-task regression symptom predictions and RMSE and CCC of 5.37 and 0.53 using the second stage RF approach, which uses single-task classification symptom predictions on E-DAIC dataset \cite{b5}.

\subsection{Large Language Models Based Approaches}

In recent years, several researchers have started using Large Language Models for depression assessment, as done in \cite{b3}, \cite{b30} and \cite{b31}. Firstly, Sadeghi et al. \cite{b3} proposed the text-based architecture for depression severity analysis on the E-DAIC dataset \cite{b5} by obtaining the textual transcripts from the OpenAI's Whisper-large model, extracted its features using OpenAI's GPT 3.5 Turbo model, fine-tuned the DepRoBERTa model and extracted its features and trained the Support Vector Regression (SVR) model. They achieved the MAE of 3.56 and 4.26 and RMSE of 5.27 and 5.36 on the development and test set, respectively. Danner et al. \cite{b30} in which they combined the training set of DAIC \cite{b10} \cite{b11}, and E-DAIC \cite{b5} datasets and train the BERT model for the depression detection task. They evaluated the model using the test set of DAIC \cite{b10} \cite{b11} dataset and combination of test set both DAIC \cite{b10} \cite{b11} and E-DAIC \cite{b5} datasets and achieved precision scores of 63 per cent and 83 per cent, recall scores of 66 per cent and 82 per cent and F1 scores of 64 per cent and 82 per cent, respectively. They also performed a direct evaluation of the GPT 3.5 model and ChatGPT 4 model using a test set of DAIC \cite{b10} \cite{b11} and achieved a precision score of 78 per cent, and 70 per cent, recall score of 79 per cent and 60 per cent and F1 score of 78 per cent and 61 per cent respectively. Finally, Hadzic et al. \cite{b31} performed a direct evaluation of Chat-GPT 3.5 and GPT 4 using a test set of DAIC \cite{b10} \cite{b11} and achieved precision, recall and F1 scores of 81 per cent, 70 per cent, and 71 per cent respectively for two-class classification. They also trained the BERT model on a training set of DAIC \cite{b10} \cite{b11} and E-DAIC \cite{b5} datasets. 

% Furthermore, to demonstrate the capabilities of large language models, A. Anand et al.\cite{b48} finetuned an LLM through instructional calibration on a dataset tailored to High School Physics and utilizing retrieval augmentation. Their finetuned retrieval-augmented model, SciPhy-RAG, demonstrates significant improvements over Vicuna-7b. 

% Also, in their other papers \cite{b49} and \cite{b50} A. Anand et al. propose MM-PhyQA, a dataset featuring high school-level multimodal physics problems and 'Mathquest' a maths dataset derived from 11th and 12th standard NCERT Mathematics textbooks respectively. These papers show their study to evaluate the performance of LLMs on these datasets, and their approach highlights the potential for enhancing LLMs' capability in textual corpora with specialized datasets and techniques.

% Moreover, using the same dataset MM-PhyQA, \cite{b51} proposed an LLM-based chatbot to answer multimodal physics multiple-choice questions (MCQs). They demonstrated that their Reinforcement Learning from Human Feedback (RLHF) and Image Captioning techniques improve the quality and accuracy of the model's responses compared to supervised fine-tuned LLMs.

Furthermore, to demonstrate the capabilities of large language models, their applications can be seen across various domains, particularly in the scientific and educational sectors, where they have demonstrated significant potential. In the scientific domain, LLMs are proving beneficial for tasks such as citation generation \cite{b48,b53,b59}, where they assist in creating precise and contextually relevant citations, and grammatical error correction \cite{b50,b51}, helping improve the fluency and accuracy of scholarly writing. These advancements have made LLMs valuable tools in enhancing the quality of scientific communication.

In the educational domain, the impact of LLMs extends to disciplines such as mathematics \cite{b54,b55,b56,b57} and physics \cite{b49,b52}, where LLMs are being used to develop innovative methodologies that enhance both teaching and learning experiences. For example, in the context of physics, retrieval augmentation and fine-tuning of LLMs have been applied to improve the accuracy of physics-related question-answering systems. Additionally, LLMs are playing a role in improving learner engagement by assessing attention in e-classrooms, with frameworks like ExCEDA \cite{b58} and ECLIPSE Dataset \cite{b60} being developed to unlock new attention paradigms. Furthermore, LLMs are contributing to the field of code generation \cite{b61}, where they help automate the creation of programming code, making it easier for students and researchers to engage in computational tasks. These applications reflect the versatility of LLMs, effectively bridging their capabilities with the specific needs of both scientific research and educational development.

\vspace{1\baselineskip}

In conclusion, the existing approaches in the literature have primarily been centred around the constraints of the AVEC 2019 challenge, and the integration of recent SOTA language models (LLMs) into their architectures needs to be addressed. This omission represents a significant gap, as modern LLMs have demonstrated remarkable natural language understanding, generation, and contextual reasoning capabilities. When effectively incorporated, these models can enhance the generalizability and efficiency of various applications, from sentiment analysis to complex predictive tasks; our research aims to bridge this divide by integrating SOTA LLMs into our proposed framework. By leveraging models such as GPT, Llama models and their successors, we aim to create a more robust, adaptable, and efficient architecture capable of handling complex linguistic nuances with minimal domain-specific adjustments.

\section{Dataset}

\begin{figure}[htbp]
\centerline{\includegraphics[width=0.3\textwidth]{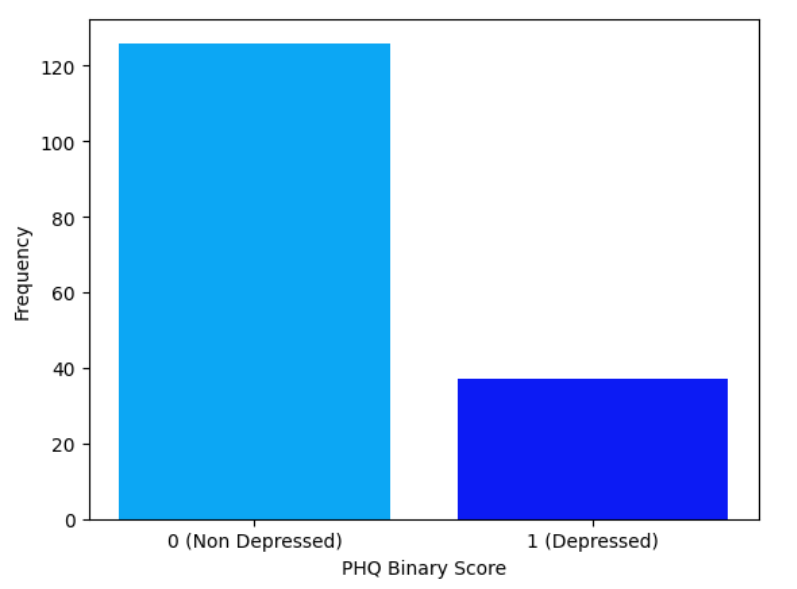}}

\caption{\small Number of Depressed \& Non-Depressed according to PHQ-8 score of Train Set}
\label{fig1}
\end{figure}

\begin{figure}[htbp]
\centerline{\includegraphics[width=0.3\textwidth]{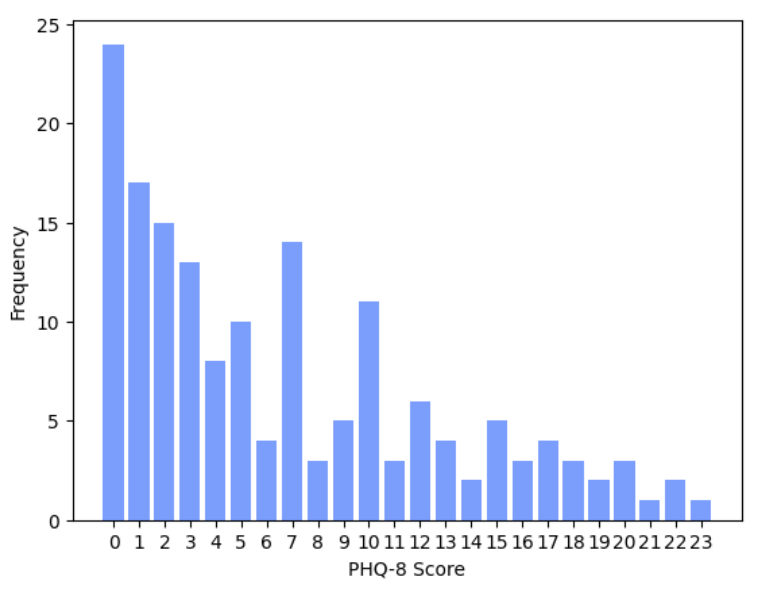}}
\caption{\small Distribution of Participants based on PHQ-8 Scores of Train Set}
\centering
\label{fig2}
\end{figure}

\begin{figure}[htbp]
\centerline{\includegraphics[width=0.3\textwidth]{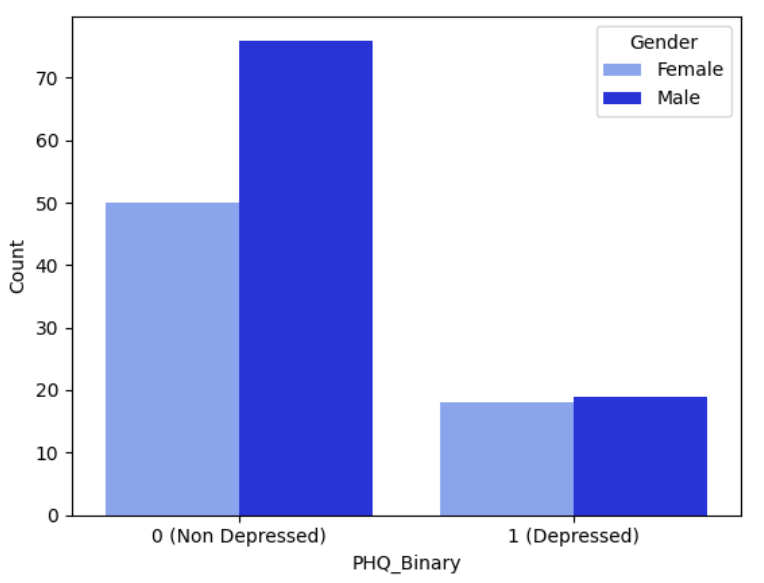}}
\caption{\small Gender Distribution Based on PHQ-8 Binary Scores of Train Set}
\label{fig3}
\end{figure}

\begin{figure}[htbp]
\centerline{\includegraphics[width=0.3\textwidth]{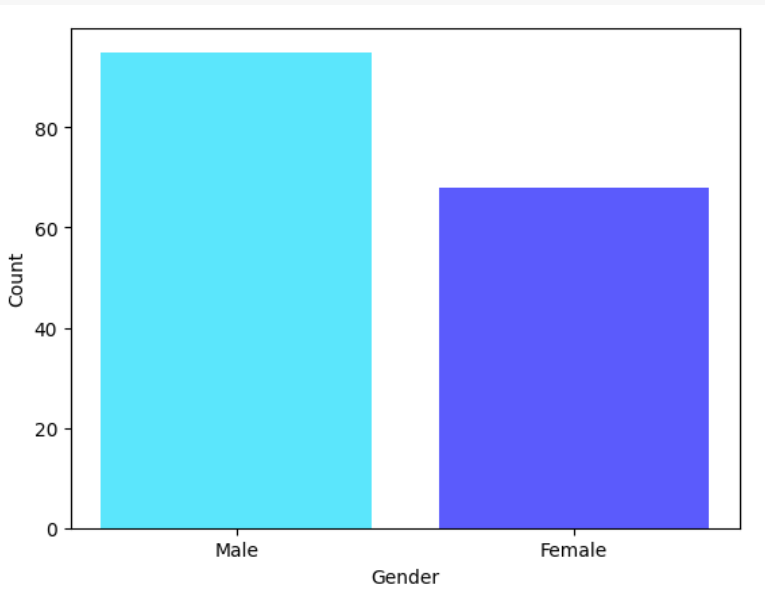}}
\caption{\small Gender Distribution of Train Set}
\label{fig4}
\end{figure}

E-DAIC \cite{b5} is an extension of the DAIC-WOZ dataset and includes semi-clinical interviews intended to assist in diagnosing psychological distress conditions, including depression, anxiety, and post-traumatic stress disorder. This dataset was also utilised in the AVEC 2019. It comprises 275 interviews conducted by an animated virtual interviewer named Ellie and by an AI-based agent, of which 163, 56, and 56 are included in the training, validation, and test sets, respectively. The training and validation sets consist of interviews taken by Ellie and AI-based agents, whereas the test set consists of only AI-based agents. Each interview record consists of the following files:
\begin{itemize}
  \item Audio files in the form of .wav format.
  \item Transcripts of the interview in the form of .csv files.
  \item Visual features obtained from the OpenFace toolkit \cite{b23}. These features include head pose features, eye gaze features, position coordinates, head rotation coordinates, and 17 Facial Action Units (FAU or AU) stored in row and column format, making it 35 features corresponding to 17 AU. These features are also in the form of .csv files.
  \item The Mel-Frequency Cepstral Coefficients (MFCCs) features extracted from the openSMILE toolkit \cite{b24}. These contain 13 MFCCs features with 13 delta and 13 double delta features combined to make 39 columns representing acoustic Low-Level Descriptors (LLD). These are in .csv file format.
  \item The eGeMaPS \cite{b43} features extracted from the openSMILE toolkit \cite{b24}. The eGeMaPS features contain 88 measures representing the LLD information of the audio file given in the dataset. These are also in .csv file format.
  \item The Bag of Video Words (BoVW) obtained from the visual features mentioned above are processed and summarised over a block of 4 seconds for each step of 1 second.
  \item The Bag of Audio Words (BoAW), derived from the above-mentioned audio features, is processed and summarised over 4-second blocks with a step duration of 1 second.
  \item Metadata files in the form .csv format consists of PHQ-8 Score, PHQ-8 Binary Score, Session ID etc. and the gender of the patient whose information is given in the table \ref{tab:phq_binary_distribution} where the PHQ Binary 0 Represents Healthy patient and one as Depressed patient and the figures \ref{fig1}, \ref{fig2}, \ref{fig3} and \ref{fig4}.
\end{itemize}

\begin{table}[ht]
    \centering
    \renewcommand{\arraystretch}{1.5} % Adjust the row height
    \begin{tabular}{|c|c|c|c|}
        \hline
        \textbf{PHQ Binary}  & \textbf{Male} & \textbf{Female} & \textbf{Total}\\
        \hline
        0 & 75 & 51 & 126 \\
        \hline
        1 & 18 & 19 & 37 \\
        \hline
        \textbf{Overall} & \textbf{93} & \textbf{70} & \textbf{163}\\
        \hline
    \end{tabular}
    \caption{\small PHQ-8 Binary Distribution by Gender of Train Set}
    \label{tab:phq_binary_distribution}
\end{table}

\section{Evaluation Metrics}

For evaluating our experiments, we have used different
evaluation metrics such as RMSE is a standard metric used to quantify the error of a model in predicting numerical data. It is used for Evaluating Regression experiments and is defined as:

\begin{equation}
\text{RMSE} = \sqrt{\frac{1}{M}\sum_{i=1}^{M} (a_i - \hat{a}_i)^2}
\end{equation}
where \( \hat{a}_i \) is the predicted value, \( a_i \) is the actual value, and \( M \) is the number of observations.

In parallel, another metric we used for regression analysis is MAE, which measures the average magnitude of errors in a set of predictions without considering their direction. It is defined as: 

\begin{equation}
\text{MAE} = \frac{1}{M} \sum_{i=1}^{M} |a_i - \hat{a}_i|
\end{equation}

where \( a_i \) is the actual value, \( \hat{a}_i \) is the predicted value, and \( M \) is the number of observations.

Finally, we also evaluated our regression results using the Concordance Correlation Coefficient (CCC) \cite{b47}, which quantifies the agreement between two variables, yielding a value between -1 and 1, with 1 indicating perfect agreement. It is defined as:

\begin{equation}
\text{CCC} = \frac{2 \eta \sigma_a \sigma_b}{\sigma_a^2 + \sigma_b^2 + (\alpha - \beta)^2}
\end{equation}

where \( \eta \) represents the Pearson correlation coefficient between the predicted and actual values, \( \sigma_a \) and \( \sigma_b \) are the standard deviations of the predicted and actual values, and \( \alpha \) and \( \beta \) are the means of the predicted and actual values.

For the classification experiments regarding the textual modality, we used accuracy, F1-scores, precision, and recall as evaluation metrics to calculate the results' performance and compare them easily with other approaches.

These metrics have been kept the same as the AVEC 2019 challenge metrics so that we can compare our results with them.

\section{Proposed Methodology}

The Methodologies adopted to analyse the various modalities, such as audio, video, and textual, are based on regression analysis across multiple models and algorithms. For the regression task, we used the PHQ-8 scores, which ranged from 0 to 24; each of the eight questions in the PHQ-8 questions was scored from 0 to 3 to determine the severity of the depression. Each score from 0-3 measures how many days the subjects have been experiencing the questionnaire problems given in the past two weeks, with 0 being "Not any day" and 3 being "Every day". Finally, the scores from these eight questions are summed to calculate the PHQ-8 score. \cite{b44}.
For the textual analysis, the methodology adopted has been based on two Supervised learning tasks: classification and regression. For classification experiments, the Subjects were divided into three classes: Healthy (PHQ-8 from 0-8), Mild (PHQ-8 from 9-15), and Depressed (PHQ-8 from 16-24). 

\subsection{Audio Data}

Upon analysis of various audio features to understand patient conditions better, we used eGeMaps features. For a comprehensive understanding, we have calculated the mean (\(\mu\)) and standard deviation (\(\sigma\)) of these features for patients divided into three classes: Healthy, Mild, and Depressive. These statistical measures help illustrate the differences and similarities in audio characteristics across the groups. The mean values indicate each feature's central tendency or average level within a group, while the standard deviation provides information about the variability or dispersion of the features around the mean. 
Table \ref{tab:comparison} presents the detailed results, showcasing \(\mu\) and \(\sigma\) for each audio feature across the classes divided based on PHQ-8 scores: PHQ-8 scores from 0-8 as Healthy, 9-15 as Mild, and scores 16-24 as depressed. This data was essential for identifying patterns and trends that can distinguish between healthy individuals and those with varying degrees of depression. By examining these metrics, we understood how certain audio features correlate with different levels of depression, ultimately aiding in more accurate and early diagnosis. \\

\begin{table*}[t]
    \centering
    \begin{tabularx}{\textwidth}{|>{\centering\arraybackslash}m{0.15\textwidth}|>{\centering\arraybackslash}X|>{\centering\arraybackslash}X|>{\centering\arraybackslash}X|>{\centering\arraybackslash}X|>{\centering\arraybackslash}X|>{\centering\arraybackslash}X|}
        \hline
        \textbf{Feature} & \multicolumn{2}{c|}{\textbf{Class Healthy}} & \multicolumn{2}{c|}{\textbf{Class Mild}} & \multicolumn{2}{c|}{\textbf{Class Depression}} \\
        \cline{2-7}
        & \textbf{Mean} & \textbf{Std} & \textbf{Mean} & \textbf{Std} & \textbf{Mean} & \textbf{Std} \\
        \hline
        Loudness & 0.091398 & 0.101953 & 0.085814 & 0.101882 & 0.074468 & 0.082629 \\
        \hline
        Hammarberg Index & 27.279813 & 8.794111 & 27.241393 & 8.768655 & 27.205442 & 8.470930 \\
        \hline
        Spectral Flux & 0.025524 & 0.043432 & 0.019987 & 0.039191 & 0.019588 & 0.035508 \\
        \hline
        Jitter & 0.005448 & 0.021401 & 0.005146 & 0.021018 & 0.004727 & 0.021474 \\
        \hline
        Shimmer & 0.255305 & 0.696044 & 0.234829 & 0.661562 & 0.211419 & 0.641799 \\
        \hline
    \end{tabularx}
    \caption{\small Comparison of audio features across different classes}
    \label{tab:comparison}
\end{table*}

As the trends Observed in \ref{tab:comparison}. Loudness Reflects the perceived intensity of the sound. Variations in loudness can indicate different emotional states. The loudness in the depression class is lower than the loudness in the non-depression class. Hammarberg Index Measures the ratio of high-frequency to low-frequency energy. It helps in assessing the quality of voice. As shown, the Hammarberg Index of depression is lower than that of the non-depression. Spectral Flux represents the rate of change in the power spectrum. It helps identify changes in speech patterns. The Spectral Flux of the depression is lower than that of the non-depression. Jitter Indicates frequency variation between cycles of the vocal waveform. High jitter values can be associated with voice pathologies. The mean Jitter of the depression is lower than that of the non-depression. Shimmer Represents amplitude variation between cycles. Like jitter, higher shimmer values can indicate vocal issues. The mean Shimmer is lower in depressed individuals than in non-depressed individuals. For Modelling tasks, we have performed the following experiments:\\ 

\subsubsection{BiLSTM}
To preprocess the audio features such that they can be used for experimentation, we clipped the first two columns ('name' and 'frame-time') from MFCC features so that only features remain; then, we normalised the MFCC values using standard scalar so that all samples are scaled. Also, to make all the samples consistent for training, we set a padding threshold of 80,000 rows. We found that MFCC features performed better than eGeMaps, so we only used MFCCs in this experiment. Although the eGeMaps did contribute to a deeper understanding of audio features as discussed in \ref{tab:comparison}.
According to the majority of approaches in the AVEC challenge 2019, BiLSTM-based networks were ubiquitous; this is due to the sequential nature of features given, which are on a frame basis for every milli second; in fact, the SOTA approach was also BiLSTM-based, so we first tried a BiLSTM-based architecture to perform regression on MFCC-feature data given in the dataset. This architecture employs an attention-based BiLSTM model. The model consists of two LSTM layers with 200 hidden units, which connect to a fully connected layer, taking 39 MFCC features as input and producing a single output for regression.\\

\subsubsection{Whisper}

Whisper \cite{b35} is an Automatic speech recognition (ASR) and Speech translation model released by OpenAI. Its remarkable speech recognition, translation, and transcribing capabilities have been leveraged for our interview audio files. We have fine-tuned the Whisper-Base model (74M parameters) applied to patients' raw interview audio files from the dataset and performed a downstream Regression task. Whisper is an encoder-decoder architecture-based model in which the encoder module transforms the input audio files into encodings, which then are given to the decoder that converts them into text tokens. So we have used the encoder module to encode the audio files, and then these 512-dimensional encodings are followed by regression layers composed of a series of fully connected layers (4098, 2048, 1024, 512, 1) with dropouts. Each layer reduces the dimensionality through linear transformations and ReLU activation, culminating in a single neuron output for performing regression. We have performed the training on a 50 GB RTX A6000 GPU for all our tasks.

\subsection{Visual Data}

For the video features, we used the Pose, Gaze, and AU pairs, represented by 49 features, containing six poses, eight gazes, and 35 AU features (which represent 17 facial action units). We have used all three of these features in combined form only as given in the dataset. For pre-processing the data, we first clipped the first four columns ('frame','timestamp','confidence', and 'success') of sample files containing the features corresponding to each patient. Then, we normalised these features using a standard scalar and truncated or padded the output files to 30,000 rows by setting a padding threshold. A similar Bi-LSTM attention-based approach was adopted for audio data analysis; the architecture includes two LSTM layers with 128 hidden units each, followed by an attention module and a single output neuron for regression output. We calculated the RMSE score on the validation set. One of the limitations of this dataset was the unavailability of the raw video files, so the experiment with pre-trained architectures was not performed. 

\subsection{Textual Data}

We have experimented with the pre-trained RoBERTa, DepRoBERTa, Proprietary models such as GPT 3.5 and GPT 4 and Open-source models such as LLAMA 3 8B instruct for textual modality. We extracted the detailed transcripts of the interviews from the respective audio file using a whisper-large-v3 model (1550 M parameters) using the full encoder-decoder architecture and then performed the task of transcription of these files to extract the text, which contains both the interviewer and the interviewee transcripts. This created a larger corpus for textual modality than was already given in the dataset. We found that these transcripts performed better than the text provided in the dataset for LLM-based experiments as they provided more context for the interviews.\\

% table
\begin{table}[ht]
\centering
\begin{tabular}{|>{\raggedright\arraybackslash}m{8cm}|}
\hline
\\\textbf{\fontsize{10}{14}\selectfont \textcolor{blue}{User Prompt:}} \\\\
\hline
\\\textbf{I have interview transcripts of many patients from a depression diagnosis interview based on PHQ-8 scores which range from 0-24, signifying 0-8 as Healthy, 9-15 as mildly depressed, 16-24 as Depressed.} \\\\
\textbf{One of the samples is following: \texttt{<Sample from train set>} The PHQ-8 score of this patient is \texttt{<score>} and in the class of \texttt{<label>}.} \\\\
\textbf{Similarly, another sample is: \texttt{<Sample from val set>} The PHQ-8 score of this patient is \texttt{<score>} and in the class of \texttt{<label>}.} \\\\
\textbf{Now predict the Exact PHQ-8 score and class of this sample: \texttt{<Sample from test set>}} \\\\
\hline
\\\textbf{\fontsize{10}{14}\selectfont \textcolor{teal}{Model Response:}} \\\\
\hline
\\\textbf{.....\texttt{<Score>}......\texttt{<Label>}.....} \\\\
\hline
\end{tabular}
\caption{\small Prompt for Predicting PHQ-8 Score and Class using GPT 3.5 and 4 Models}
\label{tab:promptgpt}
\end{table}

% table
\begin{table}[ht]
\centering
\begin{tabular}{|>{\raggedright\arraybackslash}m{8cm}|}
\hline
\\\textbf{\fontsize{10}{14}\selectfont \textcolor{blue}{System Prompt:}} \\\\
\hline
\\\textbf{Your are a depression diagnosis tool. I have interview transcripts of many patients from a depression diagnosis interview based on PHQ 8 scores which range from 0-24, signifying ( 0-8 as healthy, 9-15 as mild, 16-24 as depressed). Your task is to classify the state of depression based on the following interview transcript among the three categories and provide the exact PHQ-8 score (range 0 to 24) the patient has and nothing else. } \\\\

\hline
\\\textbf{\fontsize{10}{14}\selectfont \textcolor{red}{User Prompt:}} \\
% \hline
\\\textbf{Interview: \texttt{<sample from train set>}} \\
% \hline
\\\textbf{\fontsize{10}{14}\selectfont \textcolor{blue}{Assistant:}} \\
% \hline
\\\textbf{Label: \texttt{<label>}, score: \texttt{<PHQ-8 score>} } \\
% \hline
\\\textbf{\fontsize{10}{14}\selectfont \textcolor{red}{User Prompt:}} \\
% \hline
\\\textbf{Interview: \texttt{<sample from validation set>}} \\
% \hline
\\\textbf{\fontsize{10}{14}\selectfont \textcolor{blue}{Assistant:}} \\
% \hline
\\\textbf{Label: \texttt{<label>}, score: \texttt{<PHQ-8 score>} } \\\\
\hline
\\\textbf{\fontsize{10}{14}\selectfont \textcolor{red}{User Prompt:}} \\
% \hline
\\\textbf{Interview: \texttt{<sample from train set>}} \\\\
\hline
\\\textbf{\fontsize{10}{14}\selectfont \textcolor{teal}{Model Response:}} \\\\
\hline
\\\textbf{Label: \texttt{<label>}, score: \texttt{<PHQ-8 score>}} \\\\
\hline
\end{tabular}
\caption{\small Prompt for Predicting PHQ-8 Score and Class using Llama-3 8B}
\label{tab:promptlama}
\end{table}

\subsubsection{RoBERTa and DepRoBERTa}

RoBERTa \cite{b36} model is an extension of BERT \cite{b37} model introduced by Liu et al.  Like BERT, RoBERTa is a transformer-based language model that employs self-attention to process input sequences and generates contextualised word representations. However, RoBERTa is trained on larger datasets and employs dynamic masking during training, which helps it learn more generalisable word representations. We have used the transcripts provided in the dataset for this experiment. Firstly, we cleaned the data by expanding the acronyms as done by Ray et al. \cite{b4} and then removed the punctuations. Then, the text is encoded via the tokeniser of the DepRoBERTa model, which introduces \cite{b38}. Then, the tokenised and encoded text is trained on the RoBERTa model, and its output is used to calculate the results on a test set of the dataset. \\

\subsubsection{Proprietary LLMs GPT 3.5 and GPT 4}

GPT-3.5 \cite{b40} is an LLM developed by OpenAI comprising 175 billion parameters. It is part of the GPT models and stands for 'Generative pre-trained transformers' \cite {b45}. It utilises transformer architecture to produce text comparable to human-level proficiency based on input prompts. Trained on extensive datasets comprising vast amounts of internet text, this model can understand and generate contextually relevant and coherent responses. It can perform tasks like text completion, translation, summarisation, and conversational agents.

Whereas GPT-4 \cite{b41}, an advancement over GPT-3.5, offers enhanced text understanding, generation accuracy, and broader contextual comprehensions due to a more significant parameter estimated to be around 1.7 Trillion, which helps in more sophisticated training. This results in more reliable and nuanced outputs, making it superior for complex tasks like detailed text analysis and advanced natural language processing applications.

We have used the OpenAI's ChatGPT interface for these models as their weight checkpoints are not available as their architectures are private. We have performed two-shot prompting (as shown in Table \ref{tab:promptgpt}) where we have taken one-on-one samples from the training and validation sets, respectively, and then Prompted the LLMs to predict their PHQ-8 score and class based on given samples.\\

\subsubsection{Open-source LLM Llama-3}

Llama 3 \cite{b39} introduced by Meta AI also presented a significant advancement in open-source large language models, aiming to match the performance of leading proprietary models. Llama 3 uses a decoder-only transformer architecture. The Llama-3 8B and 70B parameter models boast a vocabulary of 128K tokens, drastically improving language encoding efficiency, consequently enhancing overall model performance set a new benchmark, demonstrating superior performance due to enhanced pretraining and post-training methods.

For using the Llama-3 model, we used the Llama-3-8B-instruct model for the prompting task. Similar to GPT models, we have performed two-shot prompting (as shown in Tables \ref{tab:promptlama}) with a slightly different prompting technique as we have not used a chat interface but followed the instruct model format to interact with the model. We have noted these output scores and classes from model output and then computed the regression and classification results. 

We have also experimented with fine-tuning llama-3-8B on our textual data, using 4-bit Quantization and LoRA techniques \cite{b46} on our 219 samples from the training and validation set To perform efficient training on available hardware. The fine-tuned model response generation suffered from hallucinations and could not provide valid responses in a specific format mentioning the PHQ-8 scores and corresponding classes in its response. This is due to fewer samples to fine-tune an 8B parameter model. An increased number of samples may be fruitful in such experiments. 

\subsection{Fusion of Modalities}

For the multimodal fusion framework, we have combined the corresponding audio and video experiments using the Whisper module and BiLSTM networks to make a novel framework that inputs audio files and video features to predict the PHQ-8 scores. We have yet to use the textual modality for this network, as the transcripts themselves are generated using the Whisper model on audio inputs, so their encodings would be the same as the audio encodings, and hence, text modality is already inherent in the whisper encodings.
The network consists of 2 Modules. The first module is the whisper encoder, which uses the whisper-base (74M) model and gives the encodings of input audio files. The second module is a BiLSTM-attention network with a configuration similar to video experimentation. These encodings are then appropriately padded for equal-sized tensors, and the output of these modules is concatenated to make encodings representing both audio and video features in the same dimensional space. These combined features are then fed to a BiLSTM-attention network of the same configuration, as the resultant features also had sequential information due to the sequential video feature embeddings. The final outputs of this BiLSTM network are given to a fully connected regression layer that predicts the PHQ-8 scores used to perform the regression experiment, as shown in figure \ref{fig:Network}.

\begin{figure}[h]
  \centering
  \includegraphics[width=0.45\textwidth]{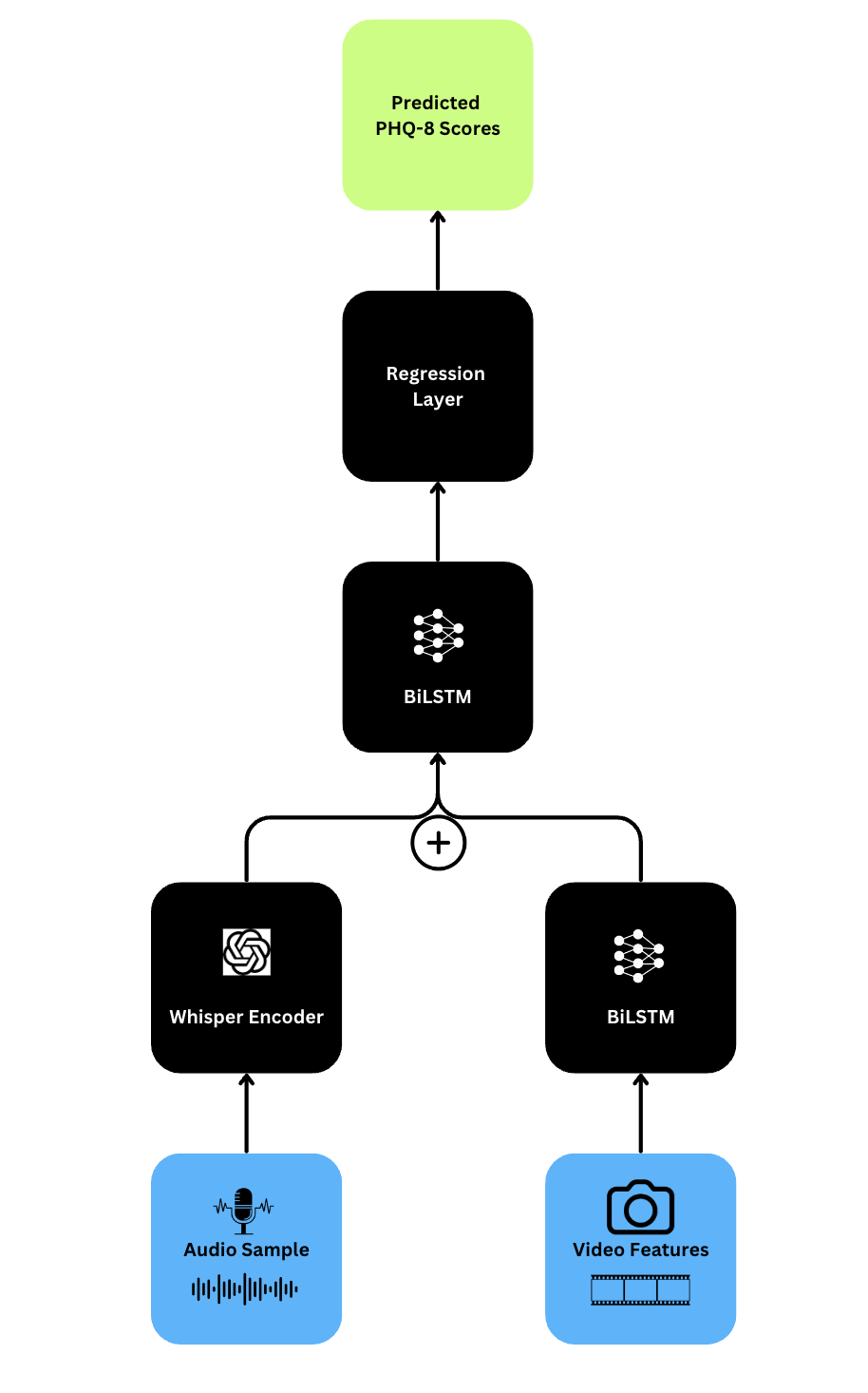}
  \caption{\small\justifying Multimodal architecture (Whisper + BiLSTM), In the proposed network, we use only the Whisper encoder to pass the encodings of audio to the next layer after being concatenated with video features to a BiLSTM network. The Video features are passed through a separate BiLSTM network before concatenation. The output of BiLSTM network is feeded to a regression layer, which outputs the Predicted PHQ-8 scores}
  \label{fig:Network}
\end{figure}

\section{Results and Discussion}
\subsection{Regression Analysis}
\subsubsection{Results on Textual Modality}

The results from different models on the textual modality in terms of evaluation metrics on the test set are displayed in Table \ref{table_1}. The textual experiments are performed on the test set in contrast to the audio and video experiments performed on the validation set. This has made it easy to compare most audio-visual and multimodal results on the validation set. The approaches on textual modalities such as in \cite{b30} have been performed in the test set, and hence, we have also adopted the evaluation on the test set. 
DeepRoBERTa tokeniser + RoBERTa model achieved the RMSE and MAE scores of 6.047 and 4.885, respectively; GPT 3.5 achieved the RMSE, MAE, and CCC scores of 5.896, 4.589 and 0.474, respectively, and LLAMA 8B Instruct model achieved the RMSE, MAE and CCC scores of 6.293, 4.893 and 0.494. On the other hand, GPT 4 achieved the RMSE, MAE and CCC scores of 3.975, 3.161 and 0.781, respectively, beating the current SOTA results for textual regression \cite{b4}. The knowledge of PHQ questionnaires helps the models accurately predict the subject's state as they try to identify the PHQ-8 answers and scores from the transcript text and aggregate them to determine the PHQ-8 score. The knowledge of class division defined in two-shot prompts also helps predict the correct score and label. \\

\begin{table}[ht]
\centering
\begin{tabular}{|c|c|c|c|}
\hline
\textbf{Model} & \textbf{RMSE} & \textbf{MAE} & \textbf{CCC}\\
\hline
DepROBERTa Tokenizer + RoBERTa Model & 6.047 & 4.885 & - \\
\hline
GPT 3.5 & 5.896 & 4.589 & 0.474 \\
\hline
GPT 4 & 3.975 & 3.161 & 0.781 \\
\hline
LLAMA 8B Instruct & 6.293 & 4.893 & 0.494 \\
\hline
\end{tabular}
\caption{\small RMSE, MAE and CCC Scores of Textual Modality on Test Set}
\label{table_1}
\end{table}

\subsubsection{Results on Audio and Visual Modalities}

The results obtained from different audio and visual modalities models on the validation set in terms of  RMSE, MAE and CCC are shown in Table \ref{table_2}. The whisper model on audio features achieves an RMSE score of 5.7 on the validation set, whereas the Bi-LSTM model on visual features achieves an RMSE of 6.45. On merging both audio and visual features, we achieved an RMSE score of 6.72 on the validation set when they were trained on the Whisper and BiLSTM model, which is 6.51 and an RMSE score of 5.39 on the validation set when trained on the BiLSTM model. 

\begin{table}[ht]
\centering
\begin{tabular}{|c|c|}
\hline
\textbf{Model} & \textbf{RMSE} \\
\hline
BiLSTM (Audio) & 5.39 \\
\hline
Whisper (Audio) & 5.7 \\
\hline
Bi-LSTM (Video) & 6.45 \\
\hline
Whisper + BiLSTM (Audio + Video) & 6.51 \\
\hline
\end{tabular}
\caption{\small RMSE Scores of Audio-Visual Modalities on Validation Set}
\label{table_2}
\end{table}

\subsubsection{Comparative Analysis}

Table \ref{tab:1} compares RMSE and CCC scores of various state-of-the-art architectures with models trained or prompt engineered on various modalities (mainly textual modality) when inferred on a test set. From Table \ref{tab:1}, it is evident that the DepRoBERTa tokeniser and RoBERTa model outperformed the baseline models cited by the competition \cite{b5} and the models proposed by Steijn et al. (using multi-task regression symptom predictions) \cite{b18}, Makiuchi et al. \cite{b20}, and Zhang et al. \cite{b29} in terms of RMSE scores. Additionally, it performed better than the model proposed by Teng et al. \cite{b19} and Zhang et al. \cite{b29} regarding MAE.

The LLAMA 8B instruct model exceeded the competition's baseline models \cite{b5} and the model by Zhang et al. \cite{b29} regarding RMSE score. It also outperforms models by Teng et al. \cite{b19} and Zhang et al. \cite{b29} in MAE and performed better than competition's baseline model \cite{b5} as well as models proposed by Fan et al. \cite{b6}, Yin et al. \cite{b12}, Teng et al. \cite{b19}, Makiuchi et al. \cite{b20}, and Saggu et al. \cite{b21} in terms of CCC score. The GPT 3.5 model, using two-shot prompting, also surpassed the competition's baseline models \cite{b5} and the models by Fan et al. \cite{b6}, Steijn et al. (with multi-task regression symptom predictions) \cite{b18}, Makiuchi et al. \cite{b20} and Zhang et al. \cite{b29} in RMSE scores while performing better than Teng et al.'s model \cite{b19}, Sun et al. \cite{b34} and Zhang et al. \cite{b28} in MAE. Additionally, it performed better than the competition's baseline model \cite{b5} as well as models proposed by Fan et al. \cite{b6}, Yin et al. \cite{b12}, Teng et al. \cite{b19}, Makiuchi et al. \cite{b20}, and Saggu et al. \cite{b21} in terms of CCC score.

Moreover, the GPT 4 model outperformed all current state-of-the-art models and the competition's baseline in RMSE, MAE, and CCC scores on the test set, establishing new SOTA performance. In audio and visual modalities, the models performed better than competition baseline models \cite{b5} but not much better than other SOTA approaches regarding RMSE scores on the validation set. 

\begin{table*}[t]
  \centering
  \begin{tabularx}{\textwidth}{|>{\centering\arraybackslash}m{0.30\textwidth}|>{\centering\arraybackslash}X|>{\centering\arraybackslash}X|>{\centering\arraybackslash}X|>{\centering\arraybackslash}X|}
    \hline 
    \textbf{Model} & \textbf{Modality} & \textbf{RMSE} & \textbf{MAE} & \textbf{CCC} \\
    \hline
    Sadeghi et al.\cite{b3} & T & 5.36 & 4.26 & - \\
    \hline
    Ray et al. \cite{b4} & T & 4.37 & 4.02 & 0.67 \\
    \hline
    Ringeval et al. (AVEC 2019 DDS Challenge Baseline Results) \cite{b5} & A, V, T & 6.37 & - & 0.111 \\
    \hline
    Fan et al.\cite{b6} & A, T & 5.91 & 4.39 & 0.43 \\
    \hline
    Yin et al.\cite{b12} & A, V, T & 5.50 & - & 0.442 \\
    \hline
    Sun et al. \cite{b13} & A, V, T & - & 4.37 & 0.583 \\
    \hline
    Yuan et al.\cite{b16} & A, V, T & 4.91 & 3.98 & 0.676 \\
    \hline
    Steijn et al. (Methodology 3) \cite{b18} & T & 6.06 & - & 0.62 \\
    \hline
    Steijn et al. (Methodology 5) \cite{b18} & T & 5.39 & - & 0.53 \\
    \hline
    Teng et al.\cite{b19} & A, V, T & - & 5.21 & 0.466 \\
    \hline
    Makiuchi et al.\cite{b20} & A, T & 6.11 & - & 0.403 \\
    \hline
    Saggu et al.\cite{b21} & A, V ,T & 5.36 & - & 0.457 \\
    \hline
    Li et al.\cite{b22} & A, V & 4.80 & 4.58 & - \\
    \hline
    Zhang et al. \cite{b28} & A  & 6.78 & 5.77 & - \\
    \hline
    Sun et al. \cite{b34} & A, V, T  & 4.31 & - & 0.491 \\
    \hline
    \textbf{DepRoBERTa Tokenizer + RoBERTa Model (Ours)} & \textbf{T} & \textbf{6.047} & \textbf{4.885} & - \\
    \hline
    \textbf{Whisper + BiLSTM (Ours)} & \textbf{A, V} & \textbf{6.51} & - & - \\
    \hline
    \textbf{GPT 3.5 (Ours)} & \textbf{T} & \textbf{5.896} & \textbf{4.589} & \textbf{0.474}\\
    \hline
    \textbf{GPT 4 (Ours)} & \textbf{T} & \textbf{3.975} & \textbf{3.161} & \textbf{0.781} \\
    \hline
    \textbf{LLAMA 8B Instruct (Ours)} & \textbf{T} & \textbf{6.293} & \textbf{4.893} & \textbf{0.494} \\
    \hline
  \end{tabularx}
  \caption{\small Comparative Analysis concerning RMSE, MAE and CCC Scores on Test Set}
  \label{tab:1}
\end{table*}

\subsection{Classification Analysis}

\begin{figure}[h]
  \centering
  \includegraphics[width=0.35\textwidth]{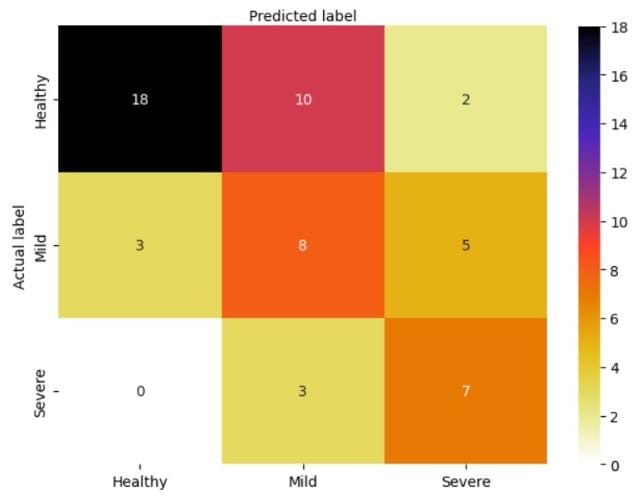}
  \caption{\small Confusion Matrix of GPT 3.5 Model on Test Set}
  \label{fig:example1}
\end{figure}

\begin{figure}[h]
  \centering
  \includegraphics[width=0.35\textwidth]{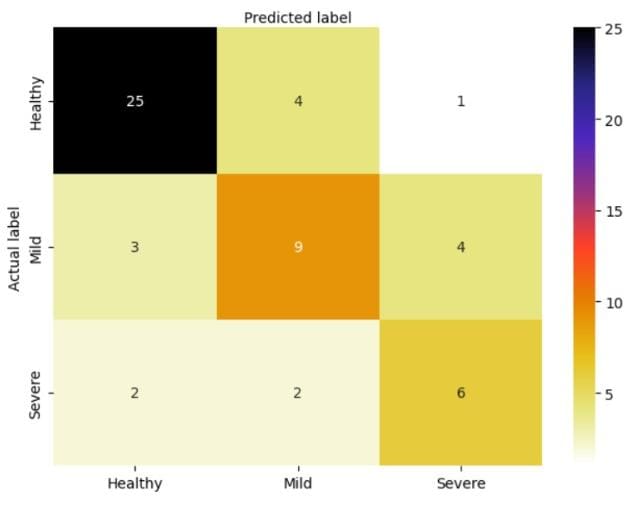}
  \caption{\small Confusion Matrix of GPT 4 Model on Test Set}
  \label{fig:example3}
\end{figure}

\begin{figure}[h]
  \centering
  \includegraphics[width=0.35\textwidth]{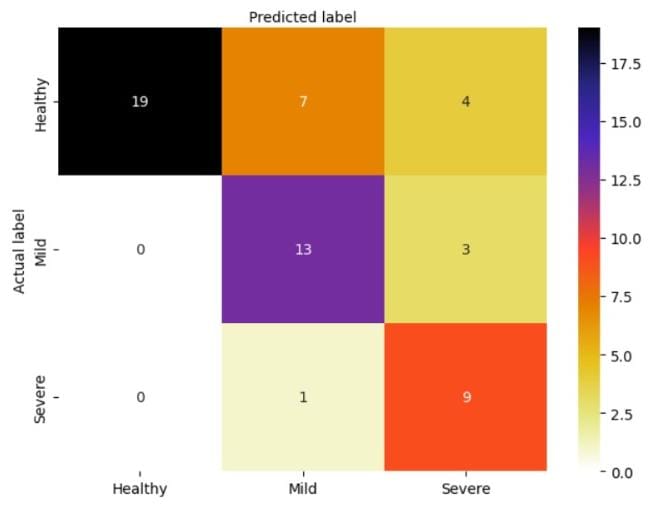}
  \caption{\small Confusion Matrix of LLAMA 3 8B Model on Test set}
  \label{fig:example5}
\end{figure}

This section discusses the classification results of all the models in textual modality. Although we tried performing classification on audio video experiments by modifying the architectures by replacing the last regressor layer with a classification layer and cross-entropy loss, the results were unsatisfactory; the confusion matrices showed all the samples classified as the healthy class.  
Therefore, the results are computed for the test set based solely on the textual modality, including metrics such as accuracy, F1 scores, precision, and recall scores. The results are in Table \ref{table_9} and the corresponding confusion matrices are in figures \ref{fig:example1} \ref{fig:example3} \ref{fig:example5}:

% \begin{itemize}

\begin{table}[t]
    \centering
    \begin{tabular}{|c|c|c|c|}
    \hline
    \textbf{Model} & \textbf{GPT 3.5} & \textbf{GPT 4} & \textbf{LLAMA 3 8B}\\
    \hline
    Accuracy & 58.93\% & 71.43\% & 73.21\% \\
    \hline
    Macro average precision & 57.94\% & 65.96\% & 72.72\% \\
    \hline
    Weighted average precision & 65.73\% & 71.53\% & 81.30\% \\
    \hline
    Macro average recall & 60.00\% & 66.53\% & 78.19\% \\
    \hline
    Weighted average recall & 58.93\% & 71.43\% & 73.21\% \\
    \hline
    Macro average F1 & 57.93\% & 66.18\% & 72.35\% \\
    \hline
    Weighted average F1 & 60.59\% & 71.44\% & 73.99\% \\
    \hline
\end{tabular}
\caption{\small Classification Analysis using GPT 3.5, GPT 4 and LLAMA 3 8B}
\label{table_9}
\end{table}

Overall, the LLAMA 3 8B model performs better than other approaches for classification tasks done via a few shot prompting on the test set, but it still needs further improvement for real-time applications. Meanwhile, GPT-4 prompting has proved to perform better for regression than the Llama 3 8B models.

\section{Conclusion and Future Scope}
In our experimentations, we have demonstrated multiple approaches on different modalities and achieved a SOTA result on regression analysis of two-shot prompting from the GPT -4 model. These Experiments show the superior capabilities of LLMs for text-based tasks such as regression and classification over other proposed architectures. 

A significant challenge during our experimentations was the limited number of samples in the dataset. The increased number of samples might result in better results by fine-tuning LLMs for textual modality. Also, raw video files and audio availability might result in more comprehensive behavioural clues for analysis and detection. The method of Data augmentation to increase the number of samples might not be ethical, given the nature and sensitivity of the data and the patient's privacy. Hence, reliable and valid data annotated by clinical experts must be created. As seen in this paper, the Textual modality, due to LLMs performance, was leading among all other modalities. For future scope, multimodal LLM-based architecture might perform better than the current SOTA networks.

\end{document}